\documentclass{ametsoc}

\journal{jas}
\bibpunct{(}{)}{;}{a}{}{,}

\title{Analytical solutions for precipitation size distributions at steady-state}

    \authors{Timothy J. Garrett\correspondingauthor{Tim Garrett, 
      135 S 1460 E Rm 819, 
     Salt Lake City, UT, 84112}}
 
     \affiliation{Department of Atmospheric Sciences, 
     University of Utah, Salt Lake City, Utah}

\email{tim.garrett@utah.edu}

\abstract{Analytical solutions are derived for the steady-state size distributions
of precipitating rain and snow particles assuming growth via collection
of suspended cloud particles. Application of the Liouville equation to the
transfer of precipitating mass through size bins in a ``cascade''  yields a characteristic
gamma distribution with a ``Marshall-Palmer'' exponential tail with
respect to particle diameter. For rain, the principle parameters controlling
size distribution shape are cloud droplet liquid water path and
cloud updraft speed. Stronger updrafts lead to greater concentrations of large
precipitating drops and a peak in the size distribution. The solutions provide a means for relating
size distributions measured in the air or on the ground to cloud bulk
microphysical and dynamic properties.}

\begin{document}

\maketitle

\section{Introduction}

In a seminal study by \citet{Marshall1948}, populations of drops
collected on dyed filter paper revealed size distributions that follow
the mathematical form $n_{D}=n_{D_{0}}\exp\left(-\lambda_{MP}D\right)$
where $n_{0}\sim8\times10^{3}$ m$^{-3}$ mm$^{-1}$, $n_{D}\Delta D$
is the concentration $n$ of rain drops of size $D$ within a bin
of width $\Delta D$, and $\lambda_{MP}=d\ln n_{D}/dD$ is the slope
parameter that would be obtained from a log-linear plot. It was shown
experimentally that $\lambda_{MP}$ (in mm$^{-1}$) can be related
to the rain rate (in mm hr$^{-1}$) through $\lambda_{MP}=4.1R^{-0.21}$. Theoretically, \citet{Villermaux2009} assumed simple
mass balance and that drops have fallspeeds of $v\propto D^{1/2}$
to obtain $\lambda\propto R^{-2/9}$, an exponent remarkably close
to the value of 0.21 obtained by \citet{Marshall1948}.
Subsequent work by \citet{GunnMarshall1958} showed a similar exponential
form for the size distributions of melted snow aggregates. 

More detailed information is provided by using a gamma distribution of form 
\begin{equation}
n_{D}=n_{0}D^{\mu}\exp\left(-\lambda D\right)\label{eq:gamma distribution}
\end{equation}
where the pre-factor exponent $\mu$ controls the numbers of smaller particle sizes. For
the purposes of representing an ensemble averaged over sufficiently
large timescales and spatial scales, this parameterized combination
of a power-law and an exponential tail provides the essential ingredients
for microwave remote sensing and cloud microphysical process algorithms
\citep{bennartz2001,MorrisonMilbrandt2015}. 

With mathematical reasoning, $\mu$ and $\lambda$ can be related
to the bulk properties of precipitation, such as the mean mass and
precipitation flux diameter \citep{Ulbrich1983,SekhonSrivastava1970,Mitchell1991}
and observations suggest that $\mu$ and $\lambda$ are nearly linearly
related \citep{McFarquhar2015}. What has yet to be fully explained
is why gamma or exponential distributions appear to describe precipitation
distributions so well, or the underlying cloud physics controlling
the values of $\mu$ and $\lambda$. The mathematical simplicity
of Eq. \ref{eq:gamma distribution} would seem to beg the question
of whether there exists an equally simple derivation. 

The most obvious approach to this problem is to explicitly simulate distribution
evolution for such key processes as vapor diffusion, collection, and
break-up \citep{Srivastava1971,List1987}. Unfortunately, an analytical
solution expressible in terms of the cloud physics at hand is not
possible without assuming \emph{a priori} some initial functional
form for the distribution \citep{feingold1988evolution}. To get a
sense for the difficulty of the problem, a single raindrop will have collected of order one million cloud droplets
of order 10 $\mu$m diameter within a timescale of order 1000 s before it reaches of a size 1 mm diameter, implying
a mean time between collisions of milliseconds. The conditions required
for initiation of this super-exponential inflation of particle volume
may be easily parameterized \citep{Garrettmodes2012}, but not the
deterministic details of what ensues. Limiting the degrees of freedom
to $n_{0}$, $\mu$ and $\lambda$ reduces complexity, but presumes
\emph{a priori} that a gamma function should apply \citep{List1987,Mitchell1991}. 

One way that a first principles equilibrium solution can be obtained is by maximizing the entropy of a particle ensemble subject to bulk constraints
on the ensemble properties. Applying this approach, \citet{Wu2018} derived a generalized form of Eq. \ref{eq:gamma distribution} that is
 $n_{D}=n_{0}D^{\mu}\exp\left(-\lambda D^{\beta}\right)$.
Unfortunately, this mathematical method does not provide explicit guidance
for the relevant ensemble constraint. For example, if the ensemble
is constrained by total precipitation mass, as might seem quite reasonable,
then $n_{D}=n_{0}D^{2}\exp\left(-\lambda D^{\beta}\right)$ where
$\beta=3$ \citep{Yano2016}. However, the observed value of $\beta$
is 1 and not 3, a difference that would greatly affect predicted concentrations
of large rain drops. 

The goal here is to derive the statistical distributions of precipitation
particles by treating the size distribution as an open system at steady-state,
defined by a continual throughput of condensed mass in a ``cascade''
through size classes. The philosophy is similar to that that taken
to derive the energy distribution of isotropic fluid turbulence with
respect to the size of the eddies, where the underlying assumptions
are only that the turbulent kinetic energy dissipation rate is independent
of eddy size \citep{TennekesandLumley} and that this energy is lost
at the smallest eddies where viscous forces balance inertial forces.
With precipitation, the cascade is of matter rather than energy, and
mass is removed gravitationally from the ensemble at all sizes and
not just at the extreme. 

\section{Generalized size distribution slope}

The start is to define a number concentration distribution of particles
with respect to mass $n_{m}=dn/dm$ such that the number concentration
of particles in a bin of width $\Delta m$ centered around $m$ is:
\begin{equation}
n\left(m-\Delta m/2,m+\Delta m/2\right)=\int_{m-\Delta m/2}^{m+\Delta m/2}n_{m}dm\label{eq:n}
\end{equation}
The Liouville equation for the evolution of $n_{m}$ in
time $t$ due to transfer along an arbitrary set of co-ordinates
$\vec{x}$ is \citep{Yano2017}: 
\begin{equation}
\frac{\partial n_{m}}{\partial t}=-\nabla\cdot\left(n_{m}\frac{d\vec{x}}{dt}\right)\label{eq:continuity}
\end{equation}
Mathematically Eq. \ref{eq:continuity} is similar to the Eulerian continuity equation. However, $\vec{x}$ does not have to be restricted to spatial
co-ordinates as it is in treatments of fluid transport of particles.
Here, it is assumed that $\vec{x}=\left(m,z\right)$ where $z$ is
a downward-pointing vertical co-ordinate, so that transport is in
and out of size bins due to the combined effects of particle growth
and vertical fallout. If horizontal spatial variability is ignored then:
\begin{equation}
\frac{\partial n_{m}}{\partial t}=-\frac{\partial}{\partial m}\left(n_{m}\frac{dm}{dt}\right)-\frac{\partial}{\partial z}\left(n_{m}\frac{dz}{dt}\right)\label{eq:continuity for precip}
\end{equation}
For particles sufficiently large that collection and precipitation
dominate, Eq. \ref{eq:continuity for precip} can be rewritten as:
\begin{equation}
\frac{\partial n_{m}}{\partial t}=\frac{\partial n_{m}}{\partial t}\bigg|_{coll}+\frac{\partial n_{m}}{\partial t}\bigg|_{precip}\label{eq:evolution of nm}
\end{equation}
So, dividing by $n_{m}$ , the dynamics can be expressed in terms
of timescales $\tau=\left|1/\left(\partial\ln n_{m}/\partial t\right)\right|$
\begin{equation}
\frac{1}{\tau}=\frac{1}{\tau_{coll}}-\frac{1}{\tau_{precip}}\label{eq:timescales}
\end{equation}
with the signs defined such that $\tau_{coll}$ is the timescale for
net transfer of particle number into a size bin of mass $m$ and
$\tau_{precip}$ is the timescale for removal. 

Because precipitation is the primary mechanism by which particles are both collected
and removed, $\tau_{coll}$ and $\tau_{precip}$ are related. Air
currents in clouds are governed by turbulent vertical motions with
their own characteristic eddy timescale $\tau_{turb}\sim2\pi/N\sim10^{3}\:{\rm s}$,
where $N\sim10^{-2}\:{\rm s^{-1}}$ is the saturated buoyancy frequency
for cloud motion adjustments with respect to their stably stratified
surroundings \citep{Durran1982,GarrettTropicalClouds2018}. Here we assume the
fallspeed of precipitation particles $v$ relative to the ground is
independent of small scale updrafts and downdrafts in the inertial
sub-range, permitting us to ignore non-equilibrium evolution over
timescales and spatial scales that are less than $\tau_{turb}$
and $\tau_{turb}v$  respectively. In this case, and provided that no precipitation falls into the cloud through its top,
the change in particle number due to settling of particles out
the bottom of a vertically and horizontally uniform cloud layer of depth $H$ at
 terminal fallspeed $v$ in an updraft of speed $w$ is: 
\begin{equation}
\frac{\partial n_{m}}{\partial t}\bigg|_{precip}=-\frac{\partial}{\partial z}\left(n_{m}\frac{dz}{dt}\right)\simeq-\frac{n_{m}\left(v-w\right)}{H}\label{eq:vertical divergence}
\end{equation}
With respect to collection, the total mass concentration flux along
the mass co-ordinate due to collection of other cloud particles is:
\begin{equation}
\frac{\partial n_{m}}{\partial t}\bigg|_{coll}=-\frac{\partial}{\partial m}\left(n_{m}\frac{dm}{dt}\right)=-\frac{\partial}{\partial m}\left(n_{m}\rho_{cloud}v\sigma\right)\label{eq:Collection}
\end{equation}
where $\rho_{cloud}=\int_{0}^{m_{c}}n_{m}dm$ is the population of
cloud particles smaller than than critical mass $m_{c}$ with
sufficiently small fall speeds to be collected by heavier particles
that have collection cross-section $\sigma$ normal to the fall velocity.
Note that collection is independent of $w$ if all particles
are buoyed equally by an updraft. 

\begin{figure}[h]
 \centerline{\includegraphics[width=20pc]{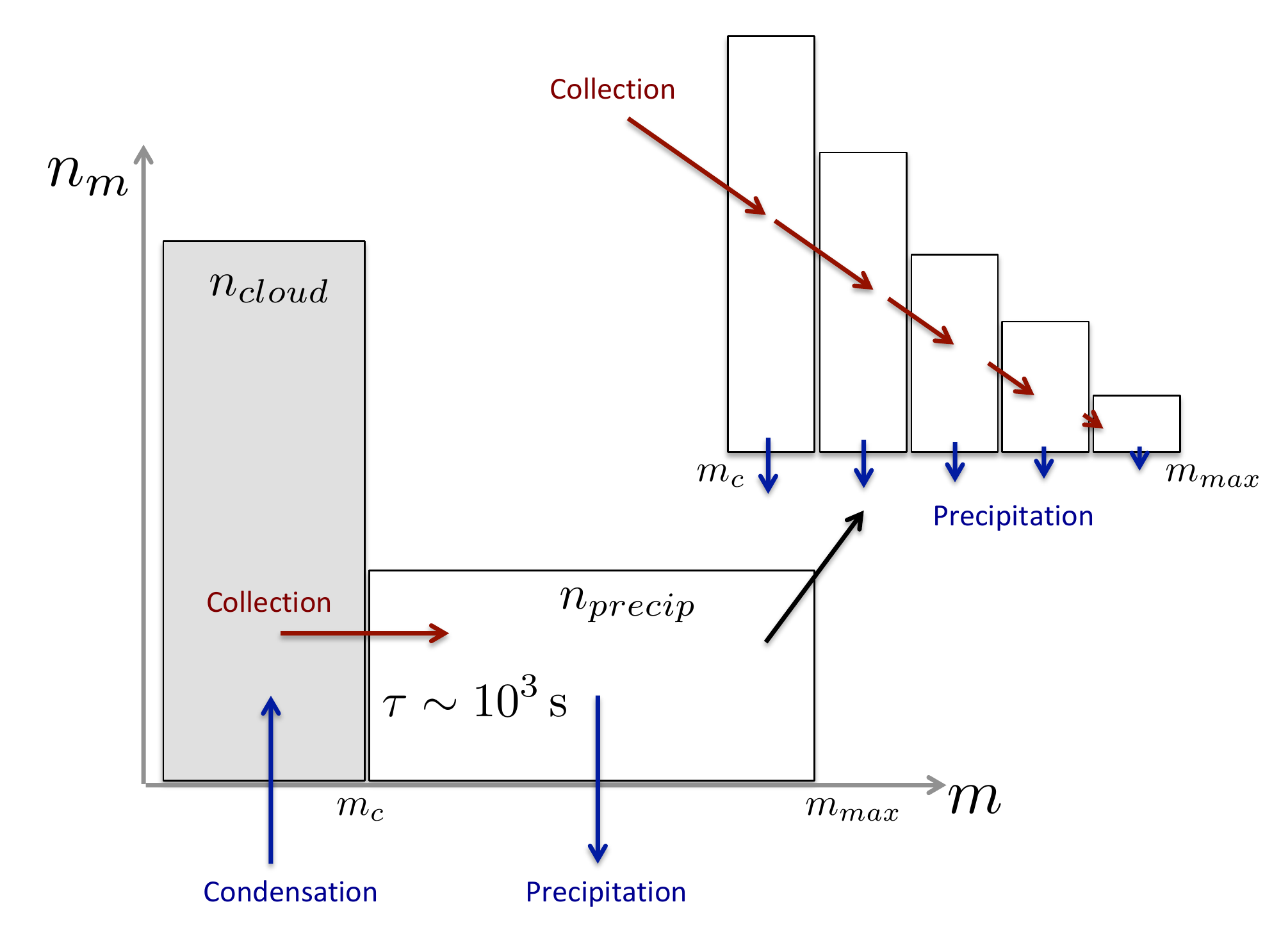}}
\caption{Illustration of precipitation distributions as an open system at steady-state,
where the time scales for condensation, collection and precipitation
are equal and net transfer of mass into precipitation size bins is
balanced by fall-out. \label{fig:Illustration-of-precipitation}}
\end{figure}

Figure \ref{fig:Illustration-of-precipitation} illustrates the steady-state
solution. For the size distribution to be stationary, such that $1/\tau\sim\partial\ln n_{m}/\partial t\simeq0$,
it follows from Eq. \ref{eq:timescales} that $\tau_{coll}\simeq\tau_{precip}$.
Dividing Eq. \ref{eq:vertical divergence} by $n_{m}$, and ignoring
for the moment updrafts, then $\tau_{precip}\sim H/v$. Very roughly,
for a small drop $\sim1$ mm diameter that falls towards the ground
at speed $v\sim1$ m s$^{-1}$ through a cloud of depth $H\sim1000$
m, the removal timescale is $\tau_{precip}\sim10^{3}$ s. From Eq.
\ref{eq:Collection}, collisions transfer mass into larger size bins
with timescale $\tau_{coll}\sim m/\left(\rho_{cloud}v\sigma\right)$.
For the the same small drop in a cloud with liquid water content $\rho_{cloud}\sim10^{-3}$
kg m$^{-3}$, $m\sim1\times10^{-6}$ kg and $\sigma\sim10^{-6}$ m$^{2}$,
then $\tau_{coll}\sim10^{3}$ s, which is similar
to $\tau_{precip}$.

A similar value of $\tau_{coll}$ is obtained for a $\sim0.1$
mm diameter drop. Moreover, the $\sim10^{3}$ s timescale  is similar
to the period of turbulent oscillations in clouds $\tau_{turb}$ that would be associated
with the production of condensate, implying a timescale
$\tau_{cond}$ that is also $\sim10^{3}$ s. Of course, crude scale analysis lacks
the details of precipitation production by clouds. Nonetheless, it serves to
illustrate an open system at steady-state defined by a continuous
throughput of condensed mass at rate $dM/dt\sim mn_{m}/\tau_{coll}$.
The rate of mass production of suspended droplets roughly equals the
rate of conversion to larger precipitating particles, and the
eventual loss of these particles by gravity.

The functional form of a stationary precipitation size distribution
can be arrived at by substituting Eqs. \ref{eq:vertical divergence} and \ref{eq:Collection} into Eq. \ref{eq:continuity for precip}:
\begin{equation}
\frac{\partial}{\partial m}\left(n_{m}\rho_{cloud} v\sigma\right)=-\frac{n_{m}\left(v-w\right)}{H}\label{eq:flux divergence}
\end{equation}
Then, dividing by Eq. \ref{eq:Collection} and noting that cloud mode particles are by definition less massive than
precipitation mode particles, in which case $m_{c}<m$ and $\partial\ln\rho_{cloud}/\partial m=0$:
\begin{equation}
\partial\ln\left(n_{m}v\sigma\right)=-\frac{\left(1-w/v\right)}{\rho_{cloud}\sigma H}\partial m\label{eq:continuity stationary}
\end{equation}
Integrating across the precipitation size distribution from $m_{c}$
to $m$: 
\begin{equation}
n_{m}v\sigma=n_{m_{c}}v_{m_{c}}\sigma_{m_{c}}\exp\left[-\frac{\int_{m_{c}}^{m}\left(1-w/v\right)dm/\sigma}{\rho_{cloud}H}\right]\label{eq:number mass distribution}
\end{equation}
For physical interpretation, Eq. \ref{eq:number mass distribution}
represents the number flux of particles through mass bins due to
collection at rate $j_{m}=n_{m}v\sigma$. With some rearranging, it can be
shown that Eq. \ref{eq:number mass distribution} is equivalent to
$j_{m}=j_{m_{c}}\exp\left(-\tau_{coll}/\tau_{precip}\right)$. As
precipitation particles grow by collection to larger values of $m$, there is a progressive decrease in the number 
of precipitation particles that remain to grow to the next size bin. Statistically
speaking, the population of particles with the longest journey has
had the greatest exposure to predation from precipitation during the period it resided in all smaller bins. If $\tau_{coll}/\tau_{precip}\gg1$,
then the particle concentration must be correspondingly small. 

Snow and rain particles can be highly deformed from sphericity \citep{Villermaux2009}.
Parameterized power-laws that provide numerical fits relating $m$
to $\sigma$ and $D$ abound in the literature \citep{Locatelli1974,Mitchell1996a}.
However, the exponent in Eq. \ref{eq:number mass distribution} must be dimensionless, imposing the
 requirement that $m/\sigma$ have units of mass per unit area. To obtain the size distribution
$n_{D}$, expressed with respect to diameter $D$, we assume the following
general forms
\begin{equation}
\sigma=\pi D^{2}/4\label{eq:area-D}
\end{equation}
\begin{equation}
m=\pi\rho_{e}D^{3}/6\label{eq:m-D}
\end{equation}
where $D$ is a cross-section effective diameter given by Eq. \ref{eq:area-D}
\citep{Locatelli1974}. The effective density $\rho_{e}$
is defined to satisfy the spherical relationship between $D$ and
$m$ for the case that $D$ refers to unmelted particles. If $\sigma$
is considered to be a collection cross-section, the efficiency of
collection by one drop of another is implicit in Eq \ref{eq:area-D}.

Substituting Eqs. \ref{eq:area-D} and \ref{eq:m-D} in Eq. \ref{eq:continuity stationary},
along with the transformation $dm=\pi\rho_{e}D^{2}dD/2$ assuming $\rho_{e}\neq\rho_{e}\left(D\right)$, the following general
solution is obtained that expresses the slope of the size distribution with
respect to $D$ on a log-linear plot:
\begin{equation}
\frac{\partial\ln n_{D}}{\partial D}= -\frac{d\ln v}{dD} - \left({1-\chi}\right){\lambda}\label{eq:slope step 1}
\end{equation}
where 
\begin{equation}
\chi=\frac{w}{v}\label{eq:chi}
\end{equation}
expresses the relative strength of the updraft velocity and
\begin{equation}
\lambda=\frac{2}{H}\frac{\rho_{e}}{\rho_{cloud}}\label{eq:Lambda}
\end{equation}
is the slope parameter that yields the Marshall-Palmer distribution
in the limit that $\chi\rightarrow0$ and $v\neq v\left(D\right)$. In terms
of the cloud equivalent liquid water path $L=\rho_{cloud}H/\rho_{l}$ where $\rho_{l}$ 
is the bulk density of liquid water:
\begin{equation}
\lambda=\frac{2}{L}\frac{\rho_{e}}{\rho_{l}}\label{eq:Lambda-L}
\end{equation}

\section{Solutions for number concentration distributions}
If the terminal velocity of a precipitation particle in still air is expressed as a power-law:
\begin{equation}
v=aD^{b}\label{eq:v-D}
\end{equation}
then Eq. \ref{eq:slope step 1} leads to:
\begin{equation}
\frac{d\ln n_{D}}{dD}=-\frac{b}{D}-(1-\frac{w}{aD^b})\lambda\label{eq:slope}
\end{equation}
A solution to Eq. \ref{eq:slope} is complicated by the functional dependence
of $b$ on $D$  \citep{Beard1976}. However, if $b$ is assumed to be a non-unity constant, then the solution mathematically resembles a gamma distribution. Integrating Eq. \ref{eq:slope} from $D_{c}\left(m_{c}\right)$ to $D$ yields 
\begin{equation}
n_{D}=n_{D_{0}}\left(\frac{D}{D_{c}}\right)^{-b}\exp\left[-\lambda D\left(1-\frac{\chi}{\left(1-b\right)}\right)\right]\label{eq:size distribution b neq 1}
\end{equation}
where $n_{D_{0}}=n_{D_{c}}\exp\left[\lambda D_{c}\left(1-\chi/\left(1-b\right)\right)\right]$. Eq. \ref{eq:size distribution b neq 1} becomes a gamma distribution for the limiting case that $\chi\rightarrow0$. A positive exponential dependence on $D$ is obtained if $w > \left(1-b\right)v$.

If $b=1$, as applies to the intermediate fallspeed regime typical of drizzle for which $0.08\:{\rm mm}<D<1.2\:{\rm mm}$, $\rho_{e}=\rho_{l}$, and $a\simeq4\times10^{3}\:{\rm s^{-1}}$, then integrating Eq. \ref{eq:slope}  from $D_{c}\left(m_{c}\right)$ to $D$ yields the gamma distribution: 
\begin{equation}
n_{D}=n_{D_{0}}\left(D/D_{c}\right)^{\mu}\exp\left(-\lambda D\right)\label{eq:size distribution general b  eq 1}
\end{equation}
where 
\begin{equation}
\mu=\lambda w/a-1=\frac{2w}{aL}-1\label{eq:nu}
\end{equation}
and $n_{D_{0}}=n_{D_{c}}\exp\left(\lambda D_{c}\right)$. Eq. \ref{eq:nu} is
qualitatively consistent with the roughly linear relationship between
$\mu$ and $\lambda$ that has been noted for arctic clouds \citep{McFarquhar2015}.
If $D_{c}$ is sufficiently close to the intercept at 0 that $\lambda D_{c}\ll1$,
or $\mu=0$, then all possible curves converge on $n_{D_{0}}\simeq n_{D_{c}}$,
consistent with observations by \citet{Marshall1948}. In the limit
$w\rightarrow0$, the power-law
pre-factor $\left(D/D_{c}\right)$ has an exponent $\mu = -1$.

If the pre-factor exponent $\mu$ in Eq. \ref{eq:nu} is positive, then there is a peak
in the gamma-distribution, requiring that $\lambda w>4\times10^{3}\:{\rm s^{-1}}$
or, by substituting Eq. \ref{eq:Lambda-L}, $w/L>2\times10^{3}\:{\rm s^{-1}}$.
In this case, the transition size $D_{t}$ representing the diameter of the peak is
obtained by setting the slope function Eq. \ref{eq:size distribution general b  eq 1}
to zero: 
\begin{equation}
D_{t}=\mu/\lambda\label{eq:Transition diameter}
\end{equation}
Substituting Eq. \ref{eq:Lambda-L} in Eq. \ref{eq:Transition diameter}
for the case that $\rho_{e}=\rho_{l}$ yields
\begin{equation}
D_{t}=w/a-L/2\label{eq:Transition diameter L}
\end{equation}
where $w$ has units of mm s$^{-1}$. 

Strictly, Eq. \ref{eq:size distribution general b  eq 1} only holds for smaller rain drops in the intermediate
fallspeed regime where $b=1$. Extending it to larger drops as a rough approximation, Eq. \ref{eq:Transition diameter L} provides a simple guide for relating the shape of the size distribution, cloud dynamics, and the cloud
water path. A precipitation size distribution characterized
by a peak with $D_{t}>0$ can be expected to be observed in measurements
of rain that falls from clouds with sufficiently high updrafts that
$w/L>2a=8\times10^{3}\:{\rm s^{-1}}$. For example, in a convective
cloud with $L>1$ mm, this would translate to precipitation forming
in an updraft of at least 8 m s$^{-1}$ with a relatively shallow
slope parameter of $\lambda>$2 mm$^{-1}$. 

An alternative simplification for integrating Eq. \ref{eq:slope step 1} is to assume a constant particle fall speed. Substituting $b=0$ into Eq. \ref{eq:size distribution b neq 1}, the solution is the Marshall-Palmer distribution:
\begin{equation}
n_{D}=n_{D_{R}}\exp\left[-\lambda_{eff}D\right]\label{eq:size distribution general b neq 1}
\end{equation}
 where
\begin{equation}
\lambda_{eff}=\left(1-\chi\right)\lambda\label{eq:lambda eff}
\end{equation}
and $\lambda$ is related to the cloud water path through Eq. \ref{eq:Lambda-L}. Eq. \ref{eq:size distribution general b neq 1}
holds in the limit of small perturbations about a reference diameter
$D_{R}$  assuming the fallspeed does not deviate significantly from
$v\left(D_{R}\right)$. A judicious reference diameter choice for
the exponential tail might be the median diameter $D_{f}$ of the
precipitation mass flux $n_{D}m\left(v-w\right)$ .

One implication of Eq. \ref{eq:lambda eff}
is that the measured size distribution depends on where it is sampled.
For example, if precipitation distributions were to be measured in
a cloud with radar, or aboard an aircraft, then it might be expected
that the effective slope parameter $\lambda_{eff}$ would be negative
for those particles with $v<w$, meaning that concentrations increase
with size. Such particles would not be expected to reach the ground. 
Only those with $v>w$ could be measured requiring that $\lambda_{eff}>0$
for the entire distribution, closer to the results obtained by
\citet{Marshall1948}. A possible counter-example is precipitation distributions
that initially formed in a strong updraft, and then only fell after
the updraft decayed. 

Effectively, the mathematical solutions provided
here assume \emph{a priori }conditions for a materially open system that is at steady-state. They 
ignore any disequilibria in fluxes in and out of size bins that might occur over timescales shorter than $\tau \sim 1000\,\rm{s}$,
or over long timescales where slow meteorological changes affect the evolution of $w$ or $H$. 
Over intermediate timescales, they are most obviously suited for stratiform precipitation, although
they could also apply to more dynamic convective precipitation provided
averaging over a sufficient amount of time and space to smooth out non-equilibrium
variability. Perturbation solutions might then be found for the case that $w$ and $H$ vary.  

\begin{figure}[h]
 \centerline{\includegraphics[width=20pc]{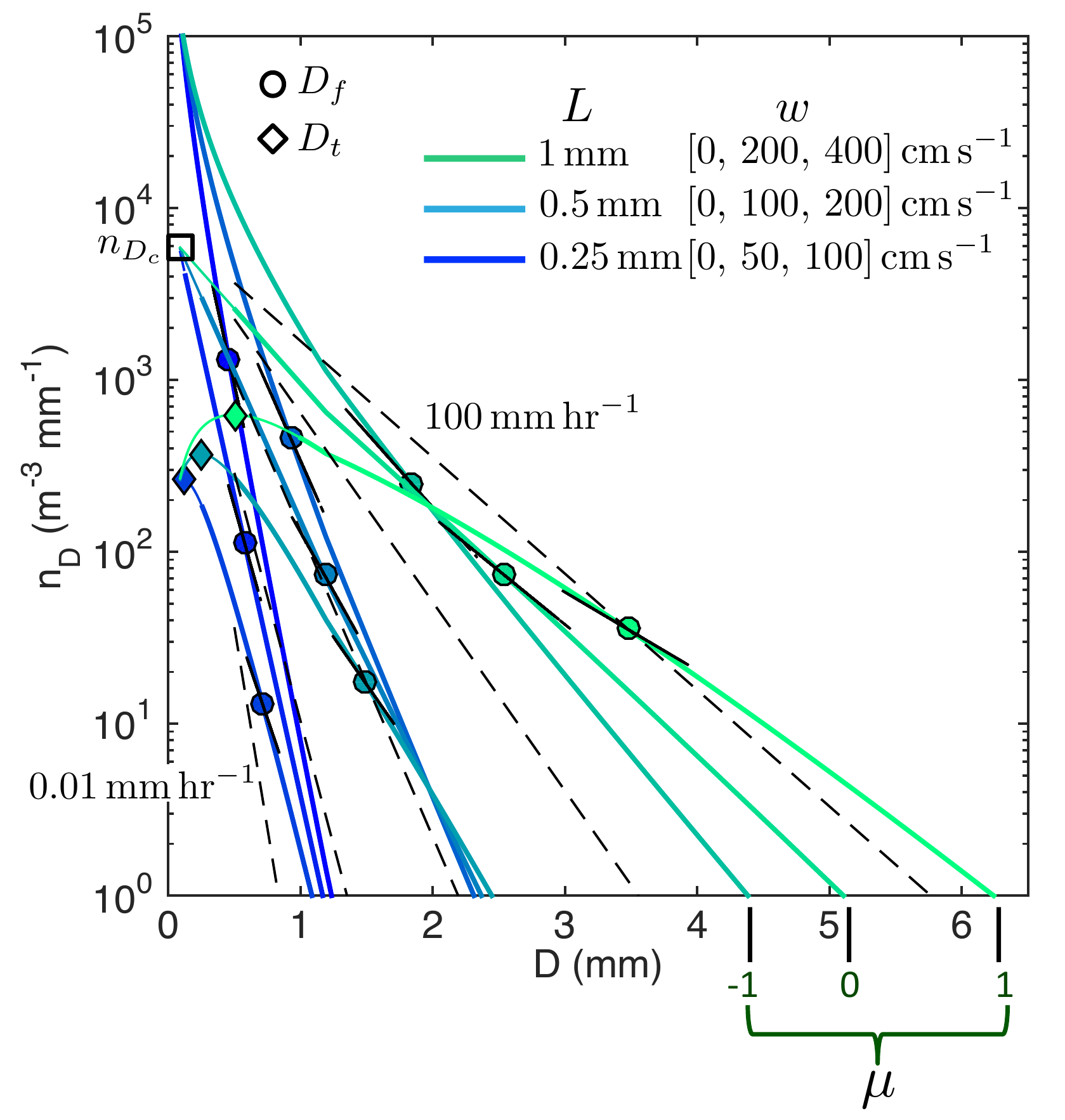}}
\caption{\label{fig:distributions}Rain size distributions obtained by integrating
Eq. \ref{eq:slope} subject to the observationally derived constraint
that $n_{D_{0}}=6\times10^{3}\exp\left(-3.2\mu\right)$ \citep{Ulbrich1983}.
Curves are grouped by three values of liquid water path $L$, and
three values of the pre-factor exponent $\mu$. The corresponding
value of $w$ is obtained from Eq. \ref{eq:nu}. Black dashed lines
represent the Marshall-Palmer distribution for rain rates ranging
from 0.01 mm hr$^{-1}$ increasing by orders of magnitude to 100 mm
hr$^{-1}$. Diamonds represent $D_{t}>0$ and circles $D_{f}$ for
each respective curve. Thin colored lines correspond to drop
sizes for which $v<w$ that would be expected not to precipitate and
thin black lines represent $\lambda_{eff}\left(D_{f}\right)$ as given
by Eq. \ref{eq:lambda eff}.}
\end{figure}

Figure \ref{fig:distributions} shows hypothetical curves for precipitation
size distributions obtained numerically by integrating Eq. \ref{eq:slope} for
the case that $v$ is a continuous function of $D$, starting from a critical diameter $D_{c}=0.04$
mm. In general, larger values of $L$ are associated with greater concentrations
of large droplets and higher rain rates. For $\mu=0$, the size distribution
approximates a pure exponential with an intercept at $n_{D_{0}}=6\times10^{3}$
\citep{Ulbrich1983}. For $\mu=-1$ corresponding with zero updraft
velocity, the size distribution is a negative power-law for small drop
sizes. High updraft scenarios with $\mu=1$ exhibit shallower
slopes in the tail of the size distribution and a peak in the distribution
that agrees well with the expression for $D_{t}$ given by Eq. \ref{eq:Transition diameter L}.
Values of $\lambda_{eff}$ for the constant fallspeed assumption $v=v\left(D_{f}\right)$
(Eq. \ref{eq:lambda eff}) match well with the local slope at $D_{f}$. High updraft speeds correspond with lower values of $\lambda_{eff}$. Values of $\lambda_{eff}$ are negative only for small
values of $D$ for which $v<w$. 

\section{Discussion}

This study did not address the impacts of turbulence on settling
speed \citep{wang1993settling,GarrettandYuter2014}, or that fragmentation
of drops might produce significant numbers of particles with super-terminal
fallspeeds \citep{Montero2009}. Fragmentation has also been invoked
as an explanation for the observed exponential tail \citep{Marshall1948,Langmuir1948}.
In support of of this hypothesis, laboratory and theoretical work
by \citet{Villermaux2009} has shown that fragmentation of large drops
$\sim6\:{\rm mm}$ exiting near cloud base can reproduce the Marshall-Palmer distribution. The fragmentation timescale of these large drops $\tau_{burst}$
is of order $10^{-2}$ s, five orders of magnitude
faster than $\tau_{coll}$. Thus, fragmentation cannot account
for a steady-state solution as there is no known cloud process that
could replenish 6 mm drops over timescales as short as $\tau_{burst}$.
Additionally, such large drops are exceedingly rare in all but
heavy precipitation. Recent photographic observations support collisions
rather than breakup as being the primary determinant of the size distribution
\citep{Testik2016,Testik2017}.

An additional limitation of the solutions described here is that they do not provide an explanation
for the value of $n_{D_{c}}\simeq n_{D_{0}}$. It was assumed here
without explicit justification that collection only applies to drops
with critical diameters $D_{c}$ greater than 0.04 mm. It is often
assumed that only drops with this size or larger have sufficiently
high collection efficiencies \citep{PruppacherKlett1997}. Alternatively, 
the 0.04 mm size threshold represents a switch
from slowing growth rates during vapor condensation to accelerating
growth rates by rapid ``discovery'' of new sources of liquid
water during collection \citep{LambVerlinde2011,Garrettmodes2012}. 

\section{Summary}

The mass continuity equation was applied to a horizontally and vertically uniform cloud layer to 
derive the size distribution of precipitation particles that exit its base, making
the equilibrium assumption that convergence of mass within a size
bin due to collection is balanced by fallout due to precipitation.
The steady-state condition yields solutions for the slope of the number
distribution with respect to particle diameter. Applying simplifying
assumptions for the dependence of fallspeed on size, the precise shape
of the distribution can be related analytically to the cloud liquid
water path $L$ and the ratio of the updraft velocity to the cloud liquid water path $w/L$. High concentrations
of large drops are related to high values of $L$ and $w/L$. If $w/L$
is sufficiently large, the local slope in the size-distribution is
positive at small drop sizes and there exists a peak in the size distribution. 

Calculations and measurements by \citet{Ulbrich1983} and \citet{Marshall1948} suggest that $R\propto\lambda_{eff}^{-4.67}$. Eq. \ref{eq:size distribution general b neq 1} for $\lambda_{eff}$ implies that rain-rates are highest coming from clouds with strong updrafts
and high liquid water paths of suspended cloud droplets. While updraft
velocities can be inferred from Doppler radar, an unfortunate limitation
of existing ground-based remote sensing techniques for measuring cloud
liquid water path is that they measure total liquid water path including
precipitation and that the sensors fail when wetted by rain \citep{Cadeddu2012}. 

Thus, the results presented here present a basis for the theoretical prediction
of relationships between precipitation microphysics, clouds and cloud
dynamics. The focus was rain, however it is straight-forward to extend
the results to snow provided the size distribution considered is with
respect to melted snow particles as in \citet{GunnMarshall1958} or
that a suitable estimate is provided for snow density $\rho_{e}$
in Eq. \ref{eq:Lambda-L} \citep{Tira2016}. Further work
is required to elucidate perturbation solutions for precipitation
events that are not in steady-state.

\acknowledgments
This work was supported by the Department of Energy Atmospheric System Research program Grant No. DE-SC0016282. The author appreciates discussions with Kyle Fitch and Chris Garrett, and comments in review from Jun-Ichi Yano and two anonymous reviewers.

 \bibliographystyle{ametsoc2014}
 \bibliography{References.bib}

\end{document}